\documentclass[conference]{IEEEtran}
\IEEEoverridecommandlockouts
\usepackage{cite}
\usepackage{amsmath,amssymb,amsfonts}
\usepackage{algorithmic}
\usepackage{graphicx}
\usepackage{textcomp}
\usepackage{xcolor}

\usepackage{multirow}
\usepackage{float}
\usepackage{array}
\usepackage{paralist} % for inline lists
\usepackage{xspace}
\usepackage{balance}
\usepackage{booktabs}
\usepackage[hidelinks]{hyperref}

\newcommand{\method}{{\scshape MulBot}\xspace}

\newcolumntype{C}[1]{>{\centering\arraybackslash}m{#1}}
\newcolumntype{L}[1]{>{\raggedright\arraybackslash}m{#1}}

\def\BibTeX{{\rm B\kern-.05em{\sc i\kern-.025em b}\kern-.08em
    T\kern-.1667em\lower.7ex\hbox{E}\kern-.125emX}}
\begin{document}

\title{\method: Unsupervised Bot Detection\\Based on Multivariate Time Series\\
}

\author{
\IEEEauthorblockN{
Lorenzo Mannocci\IEEEauthorrefmark{1}\IEEEauthorrefmark{2},
Stefano Cresci\IEEEauthorrefmark{1},
Anna Monreale\IEEEauthorrefmark{2}, 
Athina Vakali\IEEEauthorrefmark{3} and
Maurizio Tesconi\IEEEauthorrefmark{1}
}\\
\IEEEauthorblockA{\IEEEauthorrefmark{1}
Institute for Informatics and Telematics, National Research Council (IIT-CNR), Pisa, Italy \textit{[name.surname@iit.cnr.it]}}
\IEEEauthorblockA{\IEEEauthorrefmark{2}
Department of Computer Science, University of Pisa, Pisa, Italy \textit{anna.monreale@unipi.it, lorenzo.mannocci@phd.unipi.it}}
\IEEEauthorblockA{\IEEEauthorrefmark{3}
School of Informatics, Aristotle University of Thessaloniki, Thessaloniki, Greece \textit{avakali@csd.auth.gr}}
}

\maketitle

\begin{abstract}
Online social networks are actively involved in the removal of malicious social bots due to their role in the spread of low quality information. However, most of the existing bot detectors are supervised classifiers incapable of capturing the evolving behavior of sophisticated bots. Here we propose \method, an unsupervised bot detector based on multivariate time series (MTS). For the first time, we exploit multidimensional temporal features extracted from user timelines. We manage the multidimensionality with an LSTM autoencoder, which projects the MTS in a suitable latent space. Then, we perform a clustering step on this encoded representation to identify dense groups of very similar users -- a known sign of automation. Finally, we perform a binary classification task achieving f1-score $= 0.99$, outperforming state-of-the-art methods (f1-score $\le 0.97$). Not only does \method achieve excellent results in the binary classification task, but we also demonstrate its strengths in a novel and practically-relevant task: detecting and separating different botnets. In this multi-class classification task we achieve f1-score $= 0.96$. We conclude by estimating the importance of the different features used in our model and by evaluating \method's capability to generalize to new unseen bots, thus proposing a solution to the generalization deficiencies of supervised bot detectors.
\end{abstract}


\section{Introduction}
Online Social Networks (OSN) have become pervasive in our society and an ever greater source of information for many people. 
Due to their strong impact, it is paramount to address crucial issues such as disinformation, polarization and hate speech. These concerns, also considered by the European Union\footnote{\url{https://digital-strategy.ec.europa.eu/en/policies/online-disinformation}}, have a wide range of consequences, such as threatening our democracies, polarising debates, and putting the health, security, and environment of all the citizens at risk. In this scenario, social bots play an important role, especially in spreading misinformation \cite{Shao2018,Stella2018}. It was estimated that between 9\% and 17\% of Twitter accounts are bots who contribute on average between 16\% and 56\% of tweets \cite{Varol_2017,Chen_2018}. On Facebook, bots were estimated to correspond to 11\% of all accounts \cite{Cresci_2020}.
To make matters worse, many different types of bots exist with different characteristics, behaviors, and aims. For example, fake followers are the simplest kind of bots, exploited to increase the number of followers of a target account, make an account more trustworthy and influential, attracting genuine followers \cite{Cresci_2015}. We can find more complex behaviors in spambots, whose purpose is spreading (malicious) content, increasing the visibility of some public characters, and promoting a certain company's product~\cite{Cresci_2017}.

%
%Finally, accounts managed by algorithms and by human intervention are called cyborgs, defined as either bot-assisted humans or human-assisted bots \cite{chu_2012}.

The wide presence of bots in OSNs brought to the application of many machine learning models, mainly based on supervised approaches. The most widely used supervised bot detector is Botometer\footnote{\url{https://botometer.osome.iu.edu}} \cite{Davis_2016_Botometer}, which is also extensively used to label ground-truth datasets used to train other bot detectors~\cite{BotWalk}.
However, the dependency of other detectors on Botometer, which has been criticized as inaccurate~\cite{rauchfleisch2020false}, might propagate existing flaws, instead of fixing them. On the other hand, hand-labeling bot datasets does not appear to be a satisfactory solution either, given the existing biases in human labeling and the results showing that even the most experienced OSN users struggle to spot the latest generation of bots~\cite{Cresci_2017}.
%, which can be really accurate in bot detection. The most famous supervised learning model is Botometer\footnote{\url{https://botometer.osome.iu.edu}} \cite{Davis_2016_Botometer}, which is used in many works as ground truth, such as in BotWalk \cite{BotWalk}, an unsupervised approach. 
% error propagation issues, and often hand-labeling suffers from a human bias, as even the most experienced can not distinguish the latest generation of bots.

Moreover, one of the main characteristics of bots is that they \textit{evolve over time}, changing their behavior to evade detection~\cite{Cresci_2017}. This is the reason why traditional detection methods based on profile features \cite{Davis_2016_Botometer,kater_2016}, text content \cite{lee_2014,Miller_2014,DeBot_2016}, social relationships or interaction graphs \cite{jiang2016catching,jiang_2016_inferring,liu_2017} and posting patterns \cite{stringhini_2010} become soon obsolete and new techniques must be developed. There is a struggle between developers of algorithms for bot detection and increasingly sophisticated bots. The demonstration is that from 2017 onwards, there has been an increase in bot detectors based on adversarial approaches \cite{cresci_2019} as they are suitable for managing this type of task. Novel and more sophisticated bots do not act anymore individually, but as coordinated groups \cite{Cresci_2020}, forming botnets to increase their impact \cite{Zhang_2018}, such as the Star Wars botnet \cite{Echeverria_2017} or the Bursty botnet \cite{BurstyBotnet}. Supervised approaches classify individual users, thus failing to recognize coordinated bots. So, developers of bot detectors moved from supervised learning towards unsupervised models that analyze groups of accounts as a whole, rather than individual accounts, also solving the labeling issue \cite{Cresci_2020}. % , often based on individual bot detection,
In this evolving scenario, tabular features extracted from the user profile proved insufficient to identify the new bots \cite{Cresci_2020}. Indeed, considering the whole timeline of posts of the users and the synchronicity between coordinated users was shown to bring a deeper knowledge of the users' past behavior \cite{BotWalk,DNA-sequence_2018,mazza2019rtbust}.
%Temporal features of the posting content timeline of the users are generally represented as time series 

%Even in the short term (not only in the long term because of the evolutionary nature of bots), a supervised approach has significant issues. 
Finally, different types of bots (e.g., spammers, fake followers) have different behavioral characteristics. Therefore supervised learning techniques suffer a decrease in performance when trying to detect previously unseen behaviors (i.e., they do not generalize well), as also demonstrated in~\cite{Sayyadiharikandeh_2020}. The Leave One Botnet Out (LOBO) test was recently proposed to assess the generalization capabilities of bot detectors~\cite{Echeverria_2018_LOBO}. The test involves iteratively re-training bot detectors by excluding a specific botnet at each iteration and by evaluating the effectiveness at detecting bots of the excluded botnet.
%proposes an ensemble classifier, where each classifier is specialized in a different type of bot, 
%Indeed, \cite{Sayyadiharikandeh_2020} demonstrated that a classifier trained on data containing bots of a certain type fails when used to detect unseen bots. Instead, \cite{Echeverria_2018_LOBO} proposed the Leave One Botnet Out (LOBO) test to evaluate if a classifier can detect a bot class, training it only with other bot classes, explicitly without any direct knowledge of the target class itself. So, they assumed that the LOBO test is a proxy for generalization. 
%Therefore, theoretically, a characteristic of an unsupervised method is that it automatically recognizes different behaviors and clusters users according to the distribution that generated them without any lack of generalization observable in supervised approaches.

On the other hand, most unsupervised approaches rely on textual content analysis, which is now unreliable due to recent improvements in natural language generation \cite{BrownMRSKDNSSAA20,fagni_2021}. %Text features could be successful with the previous generation of bots, but current sophisticated bots produce realistic text content. 
Moreover, the few models based on temporal features, which allow tracking the whole timeline of the users, typically exploit just one dimension at a time~\cite{mazza2019rtbust}, discarding relevant information and overlooking much of the bots' behavior.

%This model allows the exploitation of user's timeline, which is useful since bots evolve over time. Moreover, it is based on multiple features, encoded as MTS, capturing complex behaviour. The unsupervised approach does not need a labelled dataset, and the clustering algorithm addresses the bot detection as a group-based approach, detecting even the most recent and sophisticated bots. This framework allows to reach  0.99 in f1-score, outperforming the state-of-the-art that achieved f1-score less or equal to 0.97. Finally, the whole approach is designed not only to distinguish bots from genuine users, but also to recognize groups of different types of bots, addressing the novel task of multi-class bots recognition. 

\subsection{Contributions}
\label{subsec:contributions}
Given the reasons mentioned above, we propose \method, an unsupervised approach for bot detection based on multivariate time series (MTS). The MTS represent temporal information of the tweet's timeline of the user. Then the multidimensional nature is addressed with the use of an LSTM autoencoder \cite{hinton_2006}. We exploit the autoencoder to compress data, reducing the original dimensionality to a smaller latent space. We choose an LSTM architecture since it is the most appropriate architecture for sequential data \cite{hochreiter_1997}. Therefore, the autoencoder encodes the MTS in a suitable latent space, where we apply the appropriate clustering algorithm. 

\method makes several contributions with respect to the state-of-the-art, as discussed in the following sections.

\paragraph{Unsupervised and Generalization} \ 
\method is completely unsupervised, overcoming the evolving nature of bots, the emergence of new types of bots and the consequent lack of generalization, typical of supervised approaches. In order to verify if \method is a solution to the lack of generalization of supervised approaches, we perform a test inspired by the LOBO test \cite{Echeverria_2018_LOBO}. We demonstrate that in a real scenario, where new kinds of bots arrive over time, the model is robust and can recognize these new ones. 

\paragraph{Multivariate Time Series}\  
\method is the first method in literature, to the best of our knowledge, based on MTS. MTS allow us to consider multiple dimensions of the users' tweets, such as the number of daily replies, hashtags, URLs and more. Previous works of the state-of-the-art, based on time series, only consider one aspect at a time, losing important information regarding bots' behavior.
\paragraph{Multi-class Classification}\  \method is the first unsupervised method that addresses multi-class bot detection. Actually, even within the supervised methods, only \cite{Dimitriadis_2021} and \cite{Sayyadiharikandeh_2020} address this task. However, both works solve the task, switching to a multiple binary classification problem. Here instead, we can separate not only bots and genuine users but also recognize different types of bots in a truly multi-class classification task. The latter is a fundamental endeavor because different types of bots have different nature, behaviors, and aims. Recognizing the kind of bot is a fundamental step in contrasting the activity of malicious bots in OSNs.
\paragraph{Results} We achieve f1-score $= 0.99$ in the traditional binary classification task, overcoming state-of-the-art methods that obtain f1-score $\le 0.97$. In the new multi-class classification task we achieve f1-score $= 0.96$. In order to make a comparison, we extend the unsupervised method by Ahmed et al.~\cite{Ahmed_2013} to the multi-class classification task, which obtains f1-score $= 0.62$.

%\textbf{Paper Organization.}
\subsection{Organization}
The rest of this paper is organized as follows.
In Section \ref{sec:related-works}, we critically discuss the current limits of supervised and unsupervised approaches and the advancement in the state-of-the-art with our approach. In Section \ref{sec:method}, we present the steps of the proposed method, \method. Section \ref{sec:experiments} describes the performed task, the experimental settings and the implementation of the method presented in Section \ref{sec:method}. In Section \ref{sec:results}, we show the reached results for each task. Lately, Section \ref{sec:conclusion} concludes the paper discussing possible future works.

\section{Related Works}
\label{sec:related-works}
In this section, we survey and critically discuss different approaches (mainly unsupervised) used in literature to perform bot detection. 

\subsection{Supervised Approaches}
We divide this section into two parts. The first one includes classic supervised machine learning approaches. While the second one presents recent deep learning methods based on neural networks.
\subsubsection{Classic Supervised Methods}
There are several works \cite{yang2020scalable,beskow2018using,beskow2018bot,beskow2018botconversations,Davis_2016_Botometer} based on features extraction and classic machine learning classifiers. The most famous supervised approach is Botometer \cite{Davis_2016_Botometer}, which is often used as ground truth for unsupervised methods. It exploits more than 1,200 features and evaluates users based on their profile information, social network, content, sentiment expressions and the timings of their actions. However, the generality and ease of deployment of this detector are counterbalanced by a reduced bot detection accuracy \cite{Cresci_2017,Grimme_2018}. In \cite{Cresci_2015}, higher performance is reached with a model specialized in a single type of bot, developing a set of supervised machine learning classifiers to detect fake followers, quite easy to identify with respect to other more sophisticated bots.
Attempting to address the evolving nature of the bots, in \cite{YangCH_2013} (2013) the authors developed a supervised classifier designed for detecting evolving bots, but yet in 2016, the proposed method was no longer successful at spotting a new wave of malicious accounts \cite{Cresci_2020}. 
In \cite{Dimitriadis_2021} an attempt to solve the generalization issue is done. Inspired by \cite{Sayyadiharikandeh_2020}, it trains an ensemble classifier, where each classifier is specialized in a type of bot. This solution only bypasses the problem, reaching a fake generalization. Indeed, a new classifier must be trained for each new type of bot, which is not an optimal solution. However, with the very similar \cite{Sayyadiharikandeh_2020}, this is the only work where a multi-class classification is addressed. Actually, in both works, the task is transformed into a multiple binary classification problem through the use of an ensemble classifier, where each internal detector classifies between genuine users and a single type of bot. So, a real multi-class classification is still not addressed. 
%Moreover, it is pretty clear how the evolving nature of bots is in contrast with the application of supervised methods. 

\subsubsection{Neural Network Methods}
In recent years, new approaches based on neural networks have been proposed with the increasing success of deep learning. In \cite{magelinski2020graph} a graph-convolutional architecture is used to extract local latent features and classify the graph based on the distribution of these features. Also \cite{alhosseini2019detect} is based on graph-convolutional networks and  \cite{feng2022heterogeneity} is always a graph-based approach, proposing a bot detector that adopts relational graph transformers to leverage the topology and heterogeneity of the real-world Twittersphere. Instead, \cite{kudugunta2018deep} leverages long short-term memory (LSTM) architecture, exploiting both content and metadata to detect bots at the tweet level. The method tries to minimize the number of features and the size of the training dataset used for classification. Also \cite{wei2019twitter} is based on text and LSTM architecture, while \cite{stanton2019gans} leverages generative adversarial networks. Finally, SATAR \cite{feng2021satar} is a self-supervised representation learning framework for Twitter users, adapting by pre-training on a massive number of self-supervised users and fine-tuning on detailed bot detection scenarios. 

\subsection{Unsupervised Approaches}
The possible unsupervised approaches can be several, based on anomaly detection or graphs that try to extract connectivity patterns of suspicious accounts or even on clustering. 
%We focus particularly on methods based on time series, extremely useful for the approach proposed in this work.

\subsubsection{Clustering and Anomaly Detection Methods}

Within works based on clustering, there are \cite{benigni2017online,wu2018bot} and \cite{koggalahewa2022unsupervised}. The latter is based on the fact that spammers do not have high overall peer acceptability and that peers would not accept them in the community. Firstly, they perform the clustering based on user interest distribution, and then they do the spam detection based on peer acceptance.
\cite{Chen_2018} detects tweets containing URLs (likely produced by spambots), with a clustering algorithm to find groups of accounts tweeting texts with high similarity. 
%A duplicate filter hashes each tweet text to a group of users and identifies groups that tweet the same text.
%It also follows the URLs in order to reconstruct the spam campaign, getting more information about malicious domain’s registrants (e.g. email).
This method's main issue is using only text features in input. In fact, it has been shown that bot detection based on text features is not a promising approach. Just think of GPT-3 \cite{BrownMRSKDNSSAA20,fagni_2021}, which can generate a text with a so high quality that it can be difficult to determine whether or not it is written by a human. 

We can find the same great limit of using text features in \cite{Miller_2014}. This work proposes a model based on two stream clustering algorithms. They address bot detection as an anomaly detection task treating spambots as outliers, relying on a vector of input features, including content, user information, and tweet text. 
%, adapting K-Means and DBSCAN respectively to stream data.
Among models based on anomaly detection also fall \cite{jiang2016catching} and \cite{RuanWWJ16}. The first one is a graph-based approach. 
%It is thought to detect compromised accounts which are genuine users that are fallen under control of an attacker.
%The method detects when a behaviour diverges significantly with respect to its associated profile. 
The drawback of the second one is that it is not a group-based approach and exploits the behavioral profile of an account, which depends only on its own actions. 

Finally, BotWalk \cite{BotWalk} is based on the fact that bots are only about 8.5\% of Twitter accounts. Therefore, it employs an ensemble anomaly detection method comparing each user to a seed-set of users (bot and normal users). The exploited features can be divided into 4 categories: metadata-, content-, temporal-, and network-based. 
%The method combines 3 different outlier scores in a unique score: Local Outlier Factor (LOF), Distance- and Angle-Based Methods, Isolation-Based method.

\subsubsection{Time Series Based Methods}
We reserve a section for unsupervised methods leveraging temporal features, which are fundamental for detecting novel sophisticated bots. Indeed, the synchronicity of the actions carried on by bots can not be bypassed by bots' developers as there must be a common purpose for the whole coordinated group of bots. So, DeBot \cite{DeBot_2016} works collecting tweets in real-time. It creates a time series for each user, with a one-second sampling rate, where zero value indicates no action and spikes represent the number of actions in that specific second. It does not distinguish different actions but sums up indiscriminately the number of actions in each second. 
Then it clusters the users hierarchically based on correlation matrix over the set of users.
%The so-called validator calculates a pairwise warped correlation matrix over the set of users and clusters the users hierarchically up to a very restrictive distance cutoff, using DTW distances and single linkage. 
This work is quite essential for our method, \method, because it is the unique case where proper time series are used even if univariate. 

Indeed, also in \cite{DNA-sequence_2018} univariate time series are exploited, but with discrete values. It creates a sequence, the so-called DNA-sequence, where each value corresponds to a user’s tweet. So the chosen granularity is the tweet, and each sequence is a succession of characters belonging to a predefined alphabet. Each character represents an entity of the tweet (hashtag, URL\ldots).
Finally, they apply an unsupervised approach based on longest common substring, which compares different sequences (users). Those who share long behavioral patterns are labeled as spambots, and users who share little similarities are labeled as genuine. If a tweet contains multiple entities, they are managed simply with \textit{tweet contains entities of mixed types}. This kind of choice implies a loss of information regarding the tweet, dealing with a model that only exploits one dimension at a time. 

A similar limit can be found in RTbust \cite{mazza2019rtbust}, which constructs a univariate time series based only on retweets. Each value represents the difference between the timestamp of the retweet and a reference timestamp. Then the time series are compressed with an LSTM variational autoencoder exploited to compress the representation and reduce the length of the time series, which all have different lengths, so to obtain a vector of the same length. It uses HDBSCAN \cite{HDBSCAN_2017} to cluster users, classifying noisy points as genuine users and clustered ones as bots.

\subsection{Advancement of the State-Of-The-Art}

\begin{table*}[t]
    \caption{Comparison of the methods sharing some properties with \method.} 
    \label{tab:comparison-models}
    \setlength{\tabcolsep}{6pt}
    \centering
    \begin{tabular}{lcC{1.3cm}C{1.5cm}C{2.2cm}C{1.5cm}C{1.8cm}cC{1cm}}
        \toprule
        \textbf{model} & \textbf{approach} & \textbf{time series} & \textbf{multiple features} & \textbf{fine-grained} & \textbf{continuous values} & \textbf{sampling granularity} & \textbf{MTS} & \textbf{multi-class} \\
        \midrule
        DeBot \cite{DeBot_2016} & unsupervised & \checkmark & \checkmark & X (sum actions) & X & 1 second & X & X\\
        DNA - Sequence \cite{DNA-sequence_2018} & unsupervised & \checkmark & \checkmark & X (``mixed type'') & X & tweet & X & X\\
        RTbust \cite{mazza2019rtbust} & unsupervised & \checkmark & X (retweet) & \checkmark & \checkmark & retweet & X & X\\
        Dimitriadis et al. \cite{Dimitriadis_2021} & supervised & X & \checkmark & \checkmark & - & - & - & \checkmark\\
        \midrule
        \method & unsupervised & \checkmark & \checkmark & \checkmark & \checkmark & 1 day & \checkmark & \checkmark\\
        \bottomrule
    \end{tabular}
\end{table*}

Starting from DeBot \cite{DeBot_2016}, DNA-sequence \cite{DNA-sequence_2018} and RTbust \cite{mazza2019rtbust}, we can highlight the properties and the deficiencies of the existing time series based model. A summary of the analysis is reported in Table \ref{tab:comparison-models}, where we can observe that none of them considers multiple features at the same time. Indeed, it is true that DeBot considers different properties with a sampling granularity of one second, but it sums up the number of actions without considering their nature. Even DNA-sequence considers different actions, but if there is more than one action per tweet (such as the presence of a mention and an URL), it assigns a symbol that means ``mixed type''. Instead, RTbust exploits the timestamp of the retweets, yet not considering other kinds of actions. Therefore, even if these methods initially include multiple features, they actually have a coarse-grained resolution in using features. We can conclude that there is an important lack in the literature, which is the use of temporal information representing the tweets’ history of the users considering several features at the same time, such as mentions, hashtags, URLs, replies and more. Finally, it is also included Dimitriadis et al. \cite{Dimitriadis_2021}, because it approaches the multi-class classification even if it is a supervised approach based on tabular features.

Therefore, \method, based on temporal information, considers the history of the user. Moreover, these multiple features represented as MTS, capturing complex behaviors, are not aggregated and are considered at a fine-grained level. The unsupervised approach is a solution to the issue of the generalization of the supervised learning method. Finally, to the best of our knowledge, it is the first unsupervised method also designed for the multi-class task, namely the recognition of the kind of bot.

\section{Method}
\label{sec:method}

\begin{figure*}[t]
    \includegraphics[width=1\linewidth]{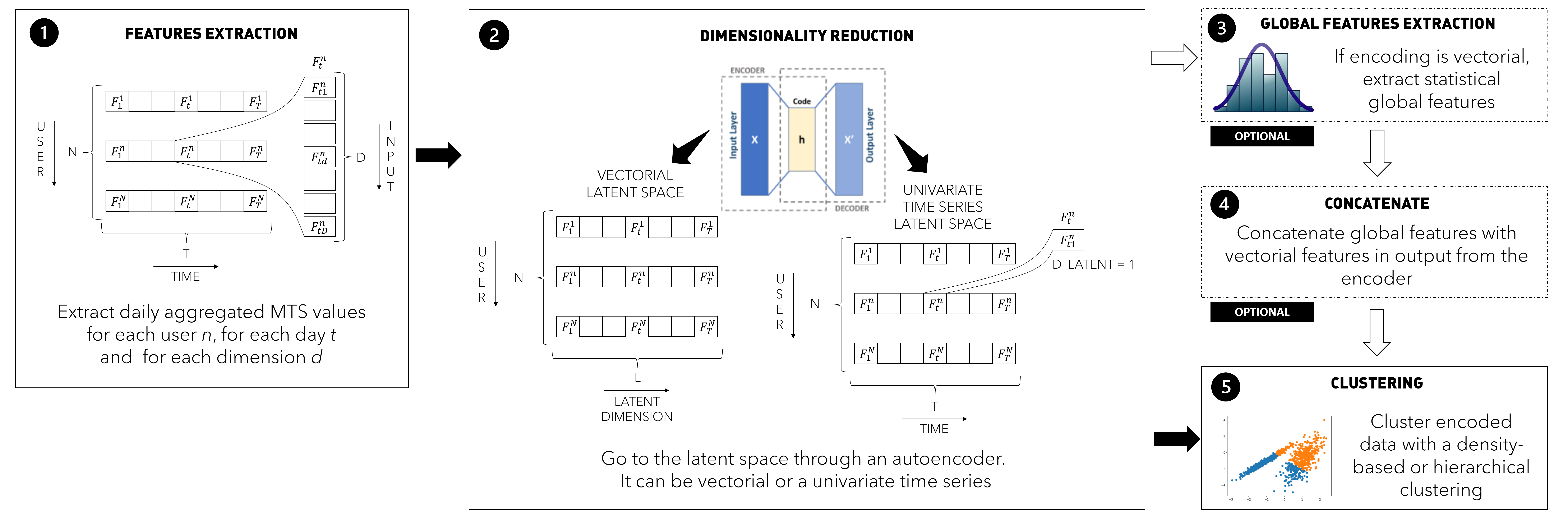}
    \centering
    \caption{\method framework, showing the  representation of the extracted MTS, the dimensionality reduction step carried on with an autoencoder, the extraction of statistical global features (optional) and finally the clustering algorithm execution.}
    % \Description{Framework of \method, showing the 5 steps of the approach. The first one is the feature extraction and the creation of the multivariate time series, which are compressed thanks to the use of an LSTM autoencoder in the second step. In the third optional step, statistical global features are extracted. In the fourth optional step, the vectorial features can be concatenated with the global statistical features. Finally clustering is performed according to the the obtained type of data.}
    \label{fig:model-framework}
\end{figure*}

We propose \method, an unsupervised approach based on multivariate time series (MTS). As discussed in the earlier sections, the MTS allow taking into account the temporal information of the users, considering more than one feature at a time. In Figure \ref{fig:model-framework}, we summarize the proposed  framework. Firstly, we extract from the tweets' timeline of the users several temporal information, aggregating them with daily granularity. This allows us to capture complex behaviors of the users and detect even sophisticated bots. Once extracted the time series, we have to manage the multidimensionality of the data. Hence, we train an autoencoder \cite{hinton_2006}, whose encoder maps the MTS into a latent space, where it is easier to perform the last step, which is the clustering algorithm. The considered latent space can have a vectorial dimensionality or can be univariate time series, leaving unaltered the temporal dimension. There is also an optional step between the dimensionality reduction of the autoencoder and the clustering algorithm. In the case of mapping toward univariate time series, we can extract global statistical features and perform the clustering on this final data. In the following sections, we describe each step of the method in detail.

\subsection{Step 1: Multivariate Time Series Extraction}
\label{subsec:time-series-extraction}
The first phase is the retrieval of information from the history of each user's tweets. We extract a set of multivariate time series describing the evolution of specific features that characterize the online user behavior. For each user, we propose to extract temporal features describing tweets, such as the number of URLs or hashtags in tweets aggregated by a daily granularity. The chosen features have already been used successfully in other forms in DNA-sequence \cite{DNA-sequence_2018} or Botwalk \cite{BotWalk}. Let $N$ be the number of users, $T$ the number of timestamps of the time series and $D$ the number of input features, the final data shaped as multivariate time series are described by a matrix $N \times T \times D$. Being the chosen granularity the day, $T$ is the number of considered days. For each user $n=1\ldots N$, for each timestamp $t=1\ldots T$, for each input dimension (i.e., feature) $d=1\ldots D$, we define the MTS value as:

\begin{equation}
    F_{t,d}^n = 
        \begin{cases}
            \sum\limits_{j=1}^{TW_t^n} f_{t,d,j}^n, & \mbox{if $TW_t^n\neq0$} \\
            -1, & \mbox{if $TW_t^n=0$}
        \end{cases}
\end{equation}

where $f_{t,d,j}^n$ is the number of occurrences of the feature $d$ (e.g. number of hashtags) in the tweet $j$, for the user $n$, in the day $t$. $TW_t^n$ is the number of tweets of the user $n$ in the day $t$. When $TW_t^n\neq0$ is the case in which there is at least a tweet in the day $t$, for the user $n$. In the case in a certain day, there are no tweets, which is a frequent situation, we assign the special value -1 to each dimension. This setting is useful for distinguishing this special case from the case in which there is a day where the user has tweets but without, for example, URLs, mentions or hashtags. In this last case, a zero value is assigned to the corresponding dimension.

\subsection{Step 2: Dimensionality Reduction}
\label{subsec:dimensionality-reduction}
Once we extracted the MTS, we have to face the issue of the multidimensional nature of the data. Inspired by RTbust \cite{mazza2019rtbust}, we train an autoencoder to use the encoder to reduce the dimensionality of the time series, mapping the MTS towards a suitable latent space. Since we are dealing with sequential data, the choice for the neural network architecture falls on an LSTM architecture \cite{hochreiter_1997}. The first possibility for the dimensionality of the latent space is the univariate time series, represented by a matrix  $N \times T \times 1$. The second one is the vectorial representation with a chosen latent dimension $L$, the encoded data have dimension $N \times L$. The encoded data is represented in step 2 of Figure \ref{fig:model-framework}.

\subsection{Step 3-4: Global features and Concatenation (optional)}
The third step is an optional phase of the method, and we can execute it only if we map the input data to the univariate time series latent space. Therefore, before the clustering step, we can extract statistical global features\footnote{An exhaustive list of the extracted features can be found: \url{https://tsfresh.readthedocs.io/en/latest/api/tsfresh.feature_extraction.html\#module-tsfresh.feature_extraction.feature_calculators}} from the univariate time series, obtaining tabular/vectorial data. Then, there is a further optional step, the fourth one. We can apply the clustering analysis to this tabular data, or we can concatenate these tabular data with the vectorial data obtained by the encoding of the MTS, using an autoencoder mapping to tabular data.

\subsection{Step 5: Clustering Analysis}
\label{subsec:clustering-analysis}
In the fifth step, data are univariate time series or vectorial features. If we map the data to the vectorial space, we can use a classical clustering algorithm for tabular data. This includes DBSCAN \cite{DBSCAN_1996}, agglomerative hierarchical clustering, HDBSCAN \cite{HDBSCAN_2017}, K-Means \cite{macqueen_1967,lloyd_1982}. HDBSCAN, already used in RTbust \cite{mazza2019rtbust}, is a mix between DBSCAN and hierarchical clustering, which are, for this reason, the most promising ones. These algorithms can also be used in case we map the original data to univariate time series. Indeed, once a similarity measure for MTS is defined, we get a matrix that can be used as input for the above algorithms.

\section{Experiments}
\label{sec:experiments}
In this section, we show the application of \method to a well-known dataset of the literature, the tasks performed, the setting of the hyperaparameters and the chosen methods of the state-of-the-art for the comparison of our results.

\subsection{Tasks}

\begin{table}[ht]
    \setlength{\tabcolsep}{6pt}
    \centering
    \caption{Support and percentage of genuine users and for each type of class of bots.}
    \label{tab:class-percentage}
    \begin{tabular}{cccrr}
        \toprule
        \textbf{class} & \textbf{bot} & \textbf{type} & \textbf{support} & \textbf{percentage}\\
        \midrule 
        0  & 0 & Genuine users & 3,394 & 29\% \\
        \cmidrule{1-5}
        1 & \multirow{4}{*}{1} & Spambots 1 & 991 & 9\% \\
        2 & & Spambots 2 & 3,457 & 30\% \\
        3 & & Spambots 3 & 464 & 4\% \\
        4 & & Fake followers & 3,202 & 28\% \\
        \bottomrule
    \end{tabular}
\end{table}

As discussed in Section \ref{sec:method}, the MTS are encoded towards a latent space, where it is possible to apply a suitable clustering algorithm. Once the data are clustered, thanks to the use of a labeled dataset, we can evaluate the results with the traditional classification metrics. Therefore, we can perform a binary and a multi-class classification in the final evaluation phase. The binary one is a state-of-the-art task whose aim is the recognition of bots from genuine users. However, here we also address a novel and more difficult task, useful for a deeper knowledge of bots' behaviors. The multi-class classification aims to divide genuine users from bots and also recognize different kind of bots. This is essential in a scenario where each kind of bots behaves in different ways, guided by different purposes \cite{Varol_2017}. 

As already mentioned, even if the method is completely unsupervised, with the use of a labeled dataset we can evaluate the results with the traditional metrics used for classification. Therefore, for the binary classification task, we report precision, recall, f1-score and accuracy (Table \ref{tab:results-binary-classification}). Here, the main considered metric is accuracy, since it is reliable when working with balanced datasets. 

Instead, in the multi-class classification task (Table \ref{tab:all-results-multi-class-classification-weighted} and Table \ref{tab:all-results-multi-class-classification-our-model}), we also consider the Matthews correlation coefficient (MCC)~\cite{chicco_2020}, since our imbalanced dataset would undermine the usefulness of the accuracy metric. %Indeed, we work in an imbalanced scenario and accuracy can be misleading.
As such, for multi-class classification we mainly consider f1-score and MCC. % , which is a more robust metric.

\begin{table}[t]
    \caption{Hyperparameters for the autoencoders with univariate time series and vecorial latent space.}
    \label{tab:hyperparameters-autoencoders}
    \centering
    \setlength{\tabcolsep}{6pt}
    \begin{tabular}{lcC{1.9cm}c}
        \toprule
        {} && \multicolumn{2}{c}{\textbf{autoencoder}}\\
        \cmidrule{3-4}
        \textbf{parameter} && \textit{univariate time series} & \textit{vecorial} \\
        \midrule 
        activation function && tanh & tanh  \\
%        \cmidrule{1-4}
        output activation function   && tanh & tanh \\
%        \cmidrule{1-4}
        optimizer && rmsprop & rmsprop \\
%        \cmidrule{1-4}
        learning rate && 0.5 & 0.0002 \\
%        \cmidrule{1-4}
        epochs && 250 & 250 \\
%        \cmidrule{1-4}
        batch size && training set size & training set size \\
%        \cmidrule{1-4}
        latent dimension && (None, 2976, 1) & 300 \\ 
        \bottomrule
    \end{tabular}
\end{table}

\subsection{Dataset}
\label{subsec:dataset}
We apply \method on a well-known Twitter dataset in bot detection literature that is cresci-17, described in \cite{Cresci_2017}.
%Even if one of the main issues in bot detection task is the lack of open datasets \cite{assenmacher_2021}, specially labeled ones, we remark that our method is entirely replicable on other datasets. 
The cresci-17 dataset includes genuine users and four kinds of bots: spambots (three types) and fake followers. In Table \ref{tab:class-percentage}, we report the support and the percentage in the dataset of genuine users and each type of bot. Overall, the dataset contains 3,394 genuine users (29\% of the total) and 8,144 bots (71\%). The large imbalance between bots and genuine users could pose challenges for the binary classification task, reason for which we balance the two classes. Instead, we keep the original (imbalanced) dataset for the multi-class task, since balancing each type of bots would result in discarding too much data.
%Therefore, the dataset is imbalanced in the percentage of genuine users/bots (29\%-71\%). This could be an issue in the binary classification, so we balance the dataset. Instead, in the multi-class scenario, it is not easy having a balanced percentage even between different classes of bots.

Therefore, for the binary classification task, to compare our approach with state-of-the-art methods mainly thought for just one kind of bot, we use the cresci-17 dataset, just including \textit{Spambots 1} and \textit{Genuine users}. Moreover, we balance the dataset with a random downsampling of the genuine users. In this way, we create the best conditions for the state-of-the-art methods for the fairest possible evaluation. Instead, in the multi-class classification task, we use the whole dataset, as we presented here.

\subsection{Features}
\label{subsec:features-input}
\begin{table}[t]
    \caption{Autoencoders structure in the univariate time series encoding and in the vectorial one.}
    \label{tab:structure-autoencoders}
    \centering
    \setlength{\tabcolsep}{6pt}
    \begin{tabular}{ll}
        \toprule
        \textbf{layer type} & \textbf{output shape} \\
        \midrule
        \multicolumn{2}{l}{\textit{LSTM Univariate Time Series}} \\ [0.7ex]
        InputLayer & (None, 2976, 6) \\
        LSTM & (None, 2976, 1) \\
        LSTM & (None, 2976, 6) \\
        \cmidrule{1-2}
        \multicolumn{2}{l}{\textit{LSTM Vectorial}} \\ [0.7ex]
        InputLayer & (None, 2976, 6) \\
        LSTM & (None, 2976, 1) \\
        Flatten & (None, 2976) \\
        Dense & (None, 300) \\
        Dense & (None, 2976) \\
        Reshape & (None, 2976, 1) \\
        LSTM & (None, 2976, 3) \\
        \bottomrule
    \end{tabular}
\end{table}
The dataset includes several user's profile features and user tweets' features, but here we use just the most effective ones according to previous works, such as DNA-sequence \cite{DNA-sequence_2018} or Botwalk \cite{BotWalk}. The features initially included in our experiments are: \begin{inparaenum}[\itshape a\upshape)]
  \item \textit{num\_urls}: number of daily URLs
  \item \textit{num\_hashtags}: number of daily hashtags
  \item \textit{num\_mentions}: number of daily mentions
  \item \textit{retweet\_count}: number of daily retweets
  \item \textit{reply\_count}: number of daily replies
  \item \textit{favorite\_count}: number of daily favorites.
\end{inparaenum}

% \begin{itemize}
%   \item \textit{num\_urls}: number of daily URLs
%   \item \textit{num\_hashtags}: number of daily hashtags
%   \item \textit{num\_mentions}: number of daily mentions
%   \item \textit{retweet\_count}: number of daily retweets
%   \item \textit{reply\_count}: number of daily reply
%   \item \textit{favorite\_count}: number of daily favorite
% \end{itemize} 

The chosen granularity is the daily aggregation, considering that a finer granularity would have caused the presence of many timestamps without tweets and a coarser one would have implied the loss of too much information. Moreover, after several preliminary trials with z-normalization and min-max normalization, the extracted MTS are normalized with min-max normalization, obtaining higher performance.

\begin{table*}[ht]
    \caption{Results for the binary classification task. In the top part, state-of-the-art methods; in the bottom part, variants of \method. Best results in each evaluation metric are shown in bold.}
    \label{tab:results-binary-classification}
    \centering
    \setlength{\tabcolsep}{6pt}
    \begin{tabular}{lcccrrrr}
        \toprule
        {} && {} && \multicolumn{4}{c}{\textbf{evaluation metrics}}\\
        \cmidrule{5-8}
        \textbf{methods} && \multicolumn{1}{c}{\textbf{type}} && \textit{precision} & \textit{recall} & \textit{f1-score}  & \textit{accuracy} \\
        \midrule 
        \multicolumn{8}{l}{\textit{State-of-the-art}} \\ [0.7ex]
        DNA - Sequence \cite{DNA-sequence_2018} && unsupervised && 0.982 & 0.972 & 0.977 & 0.976 \\
        DNA - Sequence \cite{DNA-sequence_2018} && supervised && 0.982 & 0.977 & 0.977 & 0.977 \\
        Yang et al. \cite{YangCH_2013} && supervised && 0.563 & 0.170 & 0.261 & 0.506 \\
        Miller et al. \cite{Miller_2014} && unsupervised  && 0.555 & 0.358 & 0.435 & 0.526 \\
        Ahmed\_MC && unsupervised && 0.704 & 0.699 & 0.619 & 0.698 \\ 
        Ahmed\_DBSCAN && unsupervised  && 0.930 & 0.930 & 0.930 & 0.928 \\
        \midrule
        \multicolumn{8}{l}{\textit{\method best variants}} \\ [0.7ex]
        UTS\_DBSCAN && unsupervised  && 0.855 & 0.855 & 0.855 & 0.852 \\
        UTS\_Hier  && unsupervised  && 0.935 & 0.930 & 0.925 & 0.928 \\
        Vec\_Hier && unsupervised  && 0.985 & 0.985 & \textbf{0.990} & 0.986 \\
        Glob\_Hier && unsupervised  && \textbf{0.995} & \textbf{0.995} & \textbf{0.990} & \textbf{0.993} \\
        Glob\_Vec\_Hier && unsupervised && \textbf{0.995} & \textbf{0.995} & \textbf{0.990} & 0.992 \\
        \bottomrule
    \end{tabular}
\end{table*}

\subsection{\method Implementation} 
In \method, it is required the implementation of the autoencoder and the choice of the suitable clustering algorithm.

The training of the autoencoders \cite{hinton_2006} is done with a hold-out validation and the mean squared error as loss. The chosen architecture for the neural network is the LSTM, more suitable for time series \cite{hochreiter_1997}. 
Table \ref{tab:hyperparameters-autoencoders} shows the chosen values of the hyperparameters for the two autoencoders, one mapping to the univariate time series latent space and the second one to the vectorial encoding. The first column reports the hyperparameters, while the corresponding values for the two autoencoders are reported in the second and the third columns. In table \ref{tab:structure-autoencoders}, the architecture of the two autoencoders is shown, describing the type of layer and the corresponding output dimension.

Concerning the clustering algorithm, we use for the binary classification task the Agglomerative Hierarchical algorithm with type of $linkage=Ward$\footnote{\url{https://scikit-learn.org/stable/modules/generated/sklearn.cluster.AgglomerativeClustering.html}}. Instead, for the multi-class classification, best results are obtained using a density-based clustering, namely DBSCAN \cite{DBSCAN_1996}. DBSCAN can cluster data and label specific points as noisy points\footnote{DBSCAN parameters are chosen based on the distribution of distances between points, by following the widely-used method proposed in \cite{rahmah2016determination}}. From the experiments, we observe that bots are clustered together, showing more similar behaviors with respect to genuine users, who have different behaviors. As a consequence, DBSCAN tends to consider genuine users as noisy points, unable to find a common pattern among them. This result confirms the findings presented in \cite{CresciDiPietro_2020}, where it is shown that human accounts have more heterogeneous behaviors than automated ones.
Given the discovered clusters, the first interpretation we can derive from the results is to consider all noisy points as genuine users and all the others as bots. In this case, we are simply solving a binary classification of each user. The use of a labeled dataset enables a more sophisticated inference about the type of bots represented in any discovered cluster, performing the multi-class classification task. Let $L=g, b_1, \ldots, b_k$ be the set of labels associated with the users, where $g$ is used to identify genuine users and $b_i$ represents a particular type of bot. Given the clustering result composed of a set of clusters $C_0, C_1, C_2, \ldots, C_N$, where $C_0$ represents the group of points labeled as noisy points and $C_1, C_2, \ldots, C_N$ all the other extracted clusters, we identify the users in $C_0$ as genuine while the users in each $C_i$ as bots of type $b_j$, if this label is the most frequent in the cluster $C_i$.

Lastly, we can optionally extract global statistical features with an automatic extraction process thanks to an existing package\footnote{\url{https://tsfresh.readthedocs.io/en/latest/index.html}}, from the encoded univariate time series.

\subsection{Comparisons}
\label{subsec:methods-comparisons}
For the binary classification, we compare \method against other methods of the literature, both unsupervised and supervised ones. Namely, we apply the DNA-Sequence \cite{DNA-sequence_2018} both in the unsupervised and supervised approach, Yang et al. \cite{YangCH_2013}, Miller et al. \cite{Miller_2014} and Ahmed et al. \cite{Ahmed_2013}. However, this last method expects to use the Markov Clustering, which reaches low performance in all the metrics. Therefore, we replace the Markov Clustering with DBSCAN clustering, increasing the effectiveness of the method. 

In the multi-class classification, to the best of our knowledge, there is a lack of methods with which to compare \cite{Dimitriadis_2021}. Therefore, we adapt Ahmed et al. \cite{Ahmed_2013}, implementing it also for the multi-class classification (always using DBSCAN instead of Markov Clustering). The choice of this method for this task is because it is easily extendable to the multi-class classification task, contrary to most of the methods in literature. We do not implement \cite{Sayyadiharikandeh_2020} or \cite{Dimitriadis_2021}, both for the high number of features (more than 1,200 in the first work and about 400 in the second one) and for the type of approach used: multiple binary classification tasks instead of a single multi-class task as in our work.
%Indeed, performance would be surely high, considering that a classifier is trained for each bot type in a multiple binary classification scenario.

\section{Results}
\label{sec:results}
In this section, we report the results of the tasks and experiments already presented.

\subsection{Binary Classification}
In Table \ref{tab:results-binary-classification}, we report the results for the binary classification task. In the first part of the table, we present the results of the state-of-the-art methods introduced in Section \ref{subsec:methods-comparisons}. In the second part, we report the best variants of our approach. The following tags for the variants of \method and Ahmed et al. \cite{Ahmed_2013} are used:
\begin{itemize}
  \item UTS\_DBSCAN: Encoding to univariate time series and DBSCAN as clustering algorithm;
  \item UTS\_Hier: Encoding to univariate time series and Agglomerative Hierarchical clustering algorithm, $linkage=Ward$;
  \item Vec\_Hier: Encoding to vectorial features and Agglomerative Hierarchical clustering algorithm, $linkage=Ward$;
  \item Glob\_Hier: Encoding to univariate time series, from which global features are extracted. On these global features it is performed the Agglomerative Hierarchical clustering algorithm, $linkage=Ward$; 
  \item Glob\_Vec\_Hier: Encoding to uniavriate times features, from which global features are extracted. These global features are concatenated with the vectorial ones obtained from the corresponding encoding. On this final concatenated data it is performed the Agglomerative Hierarchical clustering algorithm, $linkage=Ward$;
  \item Ahmed\_MC: Implementation of Ahmed et al. with Markov Clustering according to the original method;
  \item Ahmed\_DBSCAN: Implementation of Ahmed et al. with DBSCAN to increase the performance.
\end{itemize}
With Glob\_Hier variant, we reach the highest accuracy of 0.993. The last row of Table \ref{tab:results-binary-classification} reports the most complex case, in which the global features are concatenated with the vectorial encoding. However, there are no improvements to justify the use of this more complex variant. \method outperforms all the state-of-the-art approaches, both unsupervised and supervised.

\subsection{Multi-class Classification}
\begin{table}[t]
    \caption{Weighted performance of \method and Ahmed\_DBSCAN for multi-class classification.}
    \label{tab:all-results-multi-class-classification-weighted}
    \centering
    \setlength{\tabcolsep}{6pt}
    \begin{tabular}{crrrrr}
        \toprule
        {} & \multicolumn{5}{c}{\textbf{evaluation metrics}}\\
        \cmidrule{2-6}
        \textbf{method} & \textit{precision} & \textit{recall} & \textit{f1-score}  & \textit{MCC} & \textit{accuracy} \\
        \midrule 
        \method              & 0.966 & 0.959 & 0.963 & 0.949 & 0.966 \\
        Ahmed\_DBSCAN  & 0.704 & 0.699 & 0.618 & 0.647 & 0.698 \\
        \bottomrule
    \end{tabular}
\end{table}

\begin{table}[t]
    \caption{Results of \method for multi-class classification with input features \textit{retweet\_count}, \textit{reply\_count}, \textit{favorite\_count}, \textit{num\_mentions}, so excluding \textit{num\_urls} and \textit{num\_hashtags}.}
    \label{tab:all-results-multi-class-classification-our-model}
    \centering
    \setlength{\tabcolsep}{6pt}
    \begin{tabular}{crrrrr}
        \toprule
        {} & {} & \multicolumn{4}{c}{\textbf{evaluation metrics}}\\
        \cmidrule{3-6}
        \textbf{class} & \multicolumn{1}{c}{\textbf{support}} & \textit{precision} & \textit{recall} & \textit{f1-score}  & \textit{accuracy} \\
        \midrule 
        0  & 3,394  & 0.90 & 0.99 & 0.94 & \multirow{5}{*}{0.9624} \\
        1  & 991    & 0.99 & 0.85 & 0.92 \\
        2  & 3,457  & 1.00 & 0.98 & 0.99 \\
        3  & 464    & 1.00 & 0.82 & 0.90 \\
        4  & 3,202  & 0.99 & 0.96 & 0.98 \\
        \bottomrule
    \end{tabular}
\end{table}

\begin{figure}[t]
\includegraphics[width=85mm,scale=0.7]{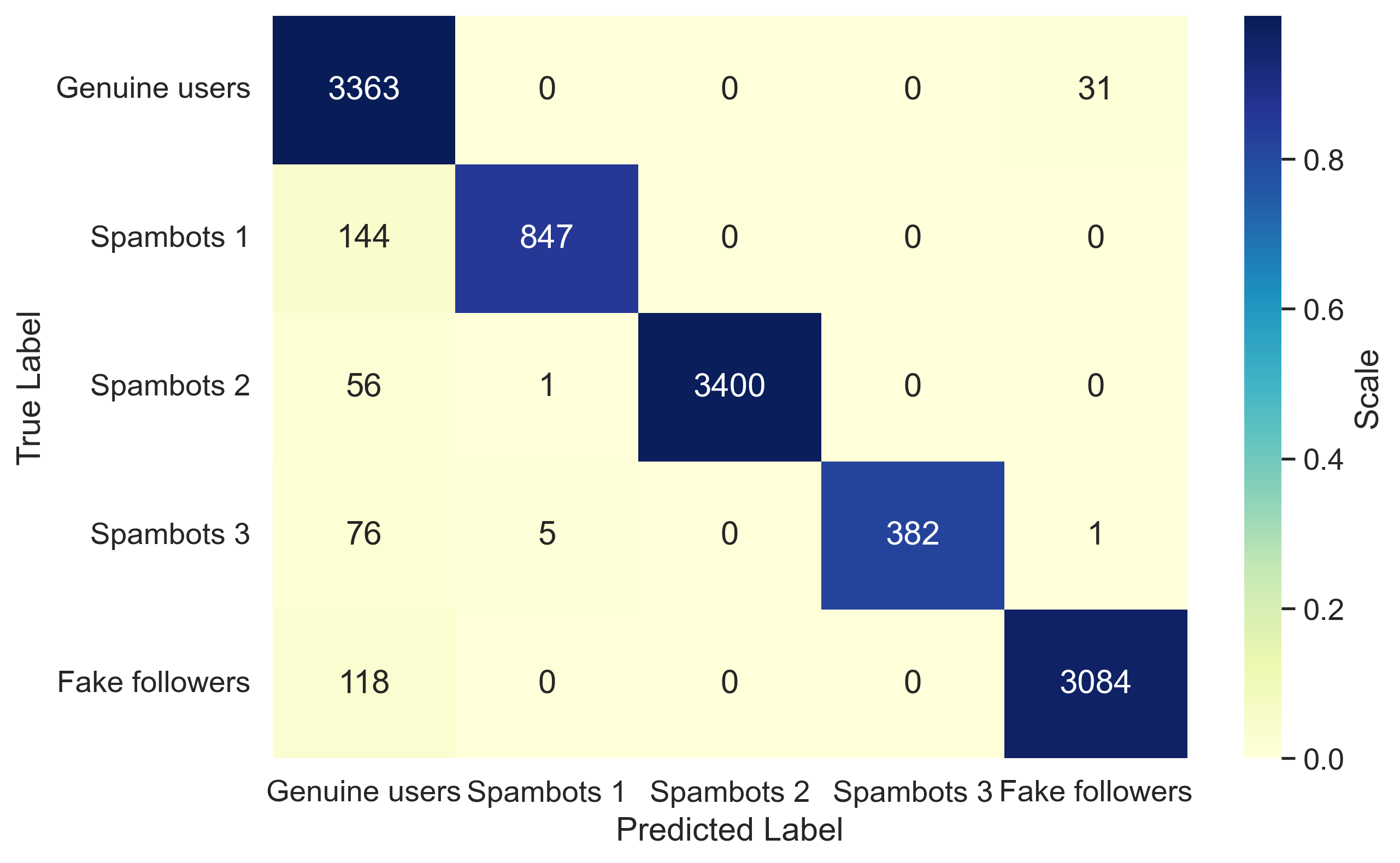}
\caption{Confusion matrix of the results of \method for multi-class classification with input features \textit{retweet\_count}, \textit{reply\_count}, \textit{favorite\_count}, \textit{num\_mentions}, so excluding \textit{num\_urls} and \textit{num\_hashtags}. The colors of the cells are based on the normalized confusion matrix, while the values in the cells are the real absolute values.} \label{fig:confusion-matrix}
% \Description{Confusion matrix for the results of \method for the multi-class classification. It shows how the method performs quite well, especially avoiding wrongly classifying genuine users as bots.}
\end{figure}

This section reports the results of the innovative task of multi-class classification in bot detection. We reach the best performance for the multi-class classification using an LSTM architecture encoding to univariate time series. Moreover the best combination of features in input includes \textit{retweet\_count}, \textit{reply\_count}, \textit{favorite\_count}, \textit{num\_mentions}, so excluding \textit{num\_urls} and \textit{num\_hash\-tags}. In Table \ref{tab:all-results-multi-class-classification-weighted} the weighted performance of the two methods is reported. We can observe that \method reaches markedly higher results both in accuracy and in f1-score with respect to Ahmed et al. \cite{Ahmed_2013}. Even if the used dataset is imbalanced, we can fairly evaluate the results by looking at the MCC (0.949), which is the most robust measure with respect to all the other ones. However, even including all the 6 features presented in Section \ref{subsec:features-input}, the method reaches 0.95 of f1-score. In Table \ref{tab:all-results-multi-class-classification-our-model}, we report the results of \method for each class, always with the best combination of features in input. For the corresponding class id, refer to Table \ref{tab:class-percentage}. 

The method reaches a precision almost equal to one for all the bots. We can better observe it also from the confusion matrix reported in Figure \ref{fig:confusion-matrix}. The colors of the cells are based on the normalized confusion matrix on a scale $[0,1]$. Instead, the values in the cells are the absolute values of the confusion matrix. Almost all the genuine users are correctly classified (row one of the confusion matrix). This means that the model does not wrongly classify genuine users as bots. This is an important result, meaning that in a real scenario, \method is careful not to “ban” genuine Twitter accounts. Instead, bots are more likely to be wrongly classified as genuine users (first column of the confusion matrix). Indeed, precision is 0.90 for \textit{Genuine users} and the recall for \textit{Spambots 1} and \textit{Spambots 3} is respectively 0.85 and 0.82. The method performs well with \textit{Fake followers}, which are one of the most simple existing bots, but also with \textit{Spambots 2}, reaching respectively an f1-score of 0.98 and 0.99.

\subsection{Feature Importance}
\label{subsec:features-importance}
\begin{figure}[t]
\includegraphics[width=85mm,scale=0.7]{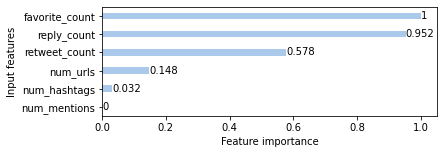}
\caption{Feature importance for the 6 features included in \method.} \label{fig:feature-importance}
\end{figure}

This section explores the feature importance of the 6 features included in \method. Inspired by \cite{Cresci_2015}, we compute a score for each feature $i$ as $S_i=\frac{f_{-i}}{f}$, where $f_{-i}$ is the f1-score obtained by \method, excluding the feature $i$ in input and $f$ is the f1-score reached by the method with all the 6 features in input. The scores are then normalized and reported in Figure \ref{fig:feature-importance}. The features \textit{favorite\_count} and \textit{reply\_count} result the most important ones. Instead, \textit{num\_mentions}, \textit{num\_hashtags}, \textit{num\_urls} have almost zero importance. These features are extracted from the tweets' content, validating that text features are not so important in bot detection tasks.

\subsection{Generalization of \method: LOBO Test}
\label{subsec:LOBO-test}
In this section, we demonstrate how \method can generalize previously unseen bots, overcoming one of the main issues of supervised approaches. With this aim, we perform the LOBO test \cite{Echeverria_2018_LOBO}. In the original work, LOBO test evaluates if a classifier can detect a bot class, training it only with other bot classes. Inspired by LOBO test, for each kind of bot $b_i$ in the dataset, we remove the $b_i$ class before the training of the autoencoder. Then we use the trained encoder to encode also the removed class $b_i$. In Table \ref{tab:LOBO-test}, we report for each experiment (one bot class removed at a time), the percentage change in the weighted f1-score, computed with respect to the \method base case that is the one reported in Table \ref{tab:all-results-multi-class-classification-weighted}. We can observe just small decreases in weighed f1-score. Therefore, our method is robust to the arrival of new types of bots and does not lack generalization, unlike supervised methods. In a real scenario, it means that the unsupervised approach continues working well, even with the arrival on the OSN of new bots with different behaviors. However, we must observe that in case of a high drop in performance, the autoencoder can be easily re-trained. Indeed, thanks to the fact that the whole method is completely unsupervised, there is no need to label the new kinds of bots, which is the main obstacle for the supervised approaches to keep up with the evolutionary nature of the bots.

\begin{table}[t]
    \caption{Percentage change of weighted f1-score for each LOBO test instance (one class removed from the training of the autoencoder at a time).}
    \label{tab:LOBO-test}
    \centering
    \setlength{\tabcolsep}{6pt}
    \begin{tabular}{crr}
        \toprule
        {} & \multicolumn{2}{c}{\textbf{evaluation metrics}}\\
        \cmidrule{2-3}
        \textbf{excluded class} & \textit{accuracy} & \textit{percentage change} \\
        \midrule 
        class 1    & 0.9577 & -0.66\% \\
        class 2    & 0.9496 & -1.39\%  \\
        class 3    & 0.9611 & -0.09\% \\
        class 4    & 0.9565 & -0.78\%  \\
        \bottomrule
    \end{tabular}
\end{table}

\subsection{Results Significance}
\method is a group-based approach, being able to recognize new sophisticated bots. The complete unsupervised nature of the method allows reactivity to the evolving nature of the bots. Moreover, the use of MTS is effective since it reaches excellent results both in binary and multi-class scenarios. This highlights how important it is the use of temporal information in the bot detection task, taking into account the whole user's timeline. The feature importance analysis confirmed the low importance of features extracted from the text, giving more importance to actions performed by other users, such as replies. Including a higher number of features in the model would allow an analysis of the features importance of a wide range of information in the bot detection task. Finally, as anticipated in Section \ref{subsec:contributions}, the developed method can reach high performance even with previously unseen bots. Therefore, it can be a solution to the issue of the generalization, typical of supervised learning approaches. 

\section{Conclusion}
\label{sec:conclusion}
We proposed an unsupervised method for bot detection to address the issues of the supervised ones. We also exploited temporal features, using multivariate time series for the first time in bot detection. We exceeded the results of the state-of-the-art methods in the binary classification task, reaching an f1-score of 0.99. We addressed for the first time a multi-classification task, achieving high results in a complex scenario where another method of the literature failed. We conducted several experiments to find the best combination of features in input, evaluating their importance and impact on the performance. Finally, we proposed the LOBO test as proxy for the generalization, showing how the proposed method can generalize to bots never seen before.

In future works, we would like to extend to multi-class classification other state-of-the-art methods to have more comparisons. We used standard features in input, but we can do a more complex features engineering, increasing the features in input and capturing more complex bots' behaviors. For example, it would be important to insert features such as the time gap between a tweet and the previous one. Moreover, being the method based on clustering, the insertion of new features could imply issues with the curse of dimensionality. Further experiments include different possibilities for the time granularity of the aggregation. The daily granularity is the most balanced in terms of MTS's length and sparsity of values. However, we could test an hourly or weekly granularity to observe the effects in the results. Finally, we could reach greater certainty of the method's effectiveness by testing it with different datasets of the literature.

\balance
\bibliographystyle{IEEEtran}
\bibliography{mybib}

\begin{thebibliography}{10}
\providecommand{\url}[1]{#1}
\csname url@samestyle\endcsname
\providecommand{\newblock}{\relax}
\providecommand{\bibinfo}[2]{#2}
\providecommand{\BIBentrySTDinterwordspacing}{\spaceskip=0pt\relax}
\providecommand{\BIBentryALTinterwordstretchfactor}{4}
\providecommand{\BIBentryALTinterwordspacing}{\spaceskip=\fontdimen2\font plus
\BIBentryALTinterwordstretchfactor\fontdimen3\font minus
  \fontdimen4\font\relax}
\providecommand{\BIBforeignlanguage}[2]{{%
\expandafter\ifx\csname l@#1\endcsname\relax
\typeout{** WARNING: IEEEtran.bst: No hyphenation pattern has been}%
\typeout{** loaded for the language `#1'. Using the pattern for}%
\typeout{** the default language instead.}%
\else
\language=\csname l@#1\endcsname
\fi
#2}}
\providecommand{\BIBdecl}{\relax}
\BIBdecl

\bibitem{Shao2018}
C.~Shao, G.~L. Ciampaglia, O.~Varol, K.-C. Yang, A.~Flammini, and F.~Menczer,
  ``The spread of low-credibility content by social bots,'' \emph{Nature
  Communications}, vol.~9, no.~1, 2018.

\bibitem{Stella2018}
M.~Stella, E.~Ferrara, and M.~D. Domenico, ``Bots increase exposure to negative
  and inflammatory content in online social systems,'' \emph{PNAS}, vol. 115,
  no.~49, 2018.

\bibitem{Varol_2017}
O.~Varol, E.~Ferrara, C.~A. Davis, F.~Menczer, and A.~Flammini, ``Online
  human-bot interactions: Detection, estimation, and characterization,'' in
  \emph{{ICWSM}}.\hskip 1em plus 0.5em minus 0.4em\relax {AAAI} Press, 2017.

\bibitem{Chen_2018}
Z.~Chen and D.~Subramanian, ``An unsupervised approach to detect spam campaigns
  that use botnets on twitter,'' 2018.

\bibitem{Cresci_2020}
S.~Cresci, ``A decade of social bot detection,'' \emph{Communications of the
  {ACM}}, vol.~63, no.~10, 2020.

\bibitem{Cresci_2015}
S.~Cresci, R.~D. Pietro, M.~Petrocchi, A.~Spognardi, and M.~Tesconi, ``Fame for
  sale: Efficient detection of fake twitter followers,'' \emph{Decis. Support
  Syst.}, vol.~80, 2015.

\bibitem{Cresci_2017}
------, ``The paradigm-shift of social spambots: Evidence, theories, and tools
  for the arms race,'' in \emph{{WWW}}.\hskip 1em plus 0.5em minus 0.4em\relax
  {ACM}, 2017.

\bibitem{Davis_2016_Botometer}
C.~A. Davis, O.~Varol, E.~Ferrara, A.~Flammini, and F.~Menczer, ``{BotOrNot}:
  {A} system to evaluate social bots,'' in \emph{{WWW} (Companion
  Volume)}.\hskip 1em plus 0.5em minus 0.4em\relax {ACM}, 2016.

\bibitem{BotWalk}
A.~J. Minnich, N.~Chavoshi, D.~Koutra, and A.~Mueen, ``Botwalk: Efficient
  adaptive exploration of {Twitter} bot networks,'' in \emph{{ASONAM}}.\hskip
  1em plus 0.5em minus 0.4em\relax {ACM}, 2017.

\bibitem{rauchfleisch2020false}
A.~Rauchfleisch and J.~Kaiser, ``The false positive problem of automatic bot
  detection in social science research,'' \emph{PLoS One}, vol.~15, no.~10,
  2020.

\bibitem{kater_2016}
C.~Kater and R.~J{\"a}schke, ``You shall not pass: detecting malicious users at
  registration time,'' in \emph{Proceedings of the 1st International Workshop
  on Online Safety, Trust and Fraud Prevention}, 2016.

\bibitem{lee_2014}
S.~Lee and J.~Kim, ``Early filtering of ephemeral malicious accounts on
  twitter,'' \emph{Computer Communications}, vol.~54, 2014.

\bibitem{Miller_2014}
Z.~Miller, B.~Dickinson, W.~Deitrick, W.~Hu, and A.~H. Wang, ``{Twitter}
  spammer detection using data stream clustering,'' \emph{Inf. Sci.}, vol. 260,
  2014.

\bibitem{DeBot_2016}
N.~Chavoshi, H.~Hamooni, and A.~Mueen, ``Debot: {Twitter} bot detection via
  warped correlation,'' in \emph{{ICDM}}.\hskip 1em plus 0.5em minus
  0.4em\relax {IEEE} Computer Society, 2016.

\bibitem{jiang2016catching}
M.~Jiang, P.~Cui, A.~Beutel, C.~Faloutsos, and S.~Yang, ``Catching synchronized
  behaviors in large networks: A graph mining approach,'' \emph{{TKDD}},
  vol.~10, no.~4, 2016.

\bibitem{jiang_2016_inferring}
------, ``Inferring lockstep behavior from connectivity pattern in large
  graphs,'' \emph{KAIS}, vol.~48, no.~2, 2016.

\bibitem{liu_2017}
S.~Liu, B.~Hooi, and C.~Faloutsos, ``Holoscope: Topology-and-spike aware fraud
  detection,'' in \emph{{CIKM}}.\hskip 1em plus 0.5em minus 0.4em\relax {ACM},
  2017.

\bibitem{stringhini_2010}
G.~Stringhini, C.~Kruegel, and G.~Vigna, ``Detecting spammers on social
  networks,'' in \emph{{ACSAC}}, 2010, pp. 1--9.

\bibitem{cresci_2019}
S.~Cresci, M.~Petrocchi, A.~Spognardi, and S.~Tognazzi, ``Better safe than
  sorry: an adversarial approach to improve social bot detection,'' in
  \emph{WebSci}, 2019.

\bibitem{Zhang_2018}
J.~Zhang, R.~Zhang, Y.~Zhang, and G.~Yan, ``The rise of social botnets: Attacks
  and countermeasures,'' \emph{Trans. Dependable Secur. Comput.}, vol.~15,
  no.~6, 2018.

\bibitem{Echeverria_2017}
J.~Echeverria and S.~Zhou, ``Discovery, retrieval, and analysis of the'star
  wars' botnet in twitter,'' in \emph{{ASONAM}}.\hskip 1em plus 0.5em minus
  0.4em\relax {ACM}, 2017.

\bibitem{BurstyBotnet}
J.~Echeverria, C.~Besel, and S.~Zhou, ``Discovery of the twitter bursty
  botnet,'' in \emph{Data Science for Cyber-Security}.\hskip 1em plus 0.5em
  minus 0.4em\relax World Scientific, 2019.

\bibitem{DNA-sequence_2018}
S.~Cresci, R.~D. Pietro, M.~Petrocchi, A.~Spognardi, and M.~Tesconi, ``Social
  fingerprinting: Detection of spambot groups through {DNA-Inspired} behavioral
  modeling,'' \emph{Trans. Dependable Secur. Comput.}, vol.~15, no.~4, 2018.

\bibitem{mazza2019rtbust}
M.~Mazza, S.~Cresci, M.~Avvenuti, W.~Quattrociocchi, and M.~Tesconi,
  ``{RTbust}: Exploiting temporal patterns for botnet detection on {Twitter},''
  in \emph{WebSci}.\hskip 1em plus 0.5em minus 0.4em\relax {ACM}, 2019.

\bibitem{Sayyadiharikandeh_2020}
M.~Sayyadiharikandeh, O.~Varol, K.~Yang, A.~Flammini, and F.~Menczer,
  ``Detection of novel social bots by ensembles of specialized classifiers,''
  in \emph{{CIKM}}.\hskip 1em plus 0.5em minus 0.4em\relax {ACM}, 2020.

\bibitem{Echeverria_2018_LOBO}
J.~Echeverr{\'{\i}}a, E.~D. Cristofaro, N.~Kourtellis, I.~Leontiadis,
  G.~Stringhini, and S.~Zhou, ``{LOBO:} evaluation of generalization
  deficiencies in {Twitter} bot classifiers,'' in \emph{{ACSAC}}.\hskip 1em
  plus 0.5em minus 0.4em\relax {ACM}, 2018.

\bibitem{BrownMRSKDNSSAA20}
T.~Brown \emph{et~al.}, ``Language models are few-shot learners,'' in
  \emph{NEURIPS}, vol.~33.\hskip 1em plus 0.5em minus 0.4em\relax Curran
  Associates, Inc., 2020.

\bibitem{fagni_2021}
T.~Fagni, F.~Falchi, M.~Gambini, A.~Martella, and M.~Tesconi, ``Tweepfake:
  About detecting deepfake tweets,'' \emph{PloS one}, vol.~16, no.~5, 2021.

\bibitem{hinton_2006}
G.~E. Hinton and R.~R. Salakhutdinov, ``Reducing the dimensionality of data
  with neural networks,'' \emph{Science}, vol. 313, no. 5786, 2006.

\bibitem{hochreiter_1997}
S.~Hochreiter and J.~Schmidhuber, ``Long short-term memory,'' \emph{Neural
  computation}, vol.~9, no.~8, 1997.

\bibitem{Dimitriadis_2021}
I.~Dimitriadis, K.~Georgiou, and A.~Vakali, ``Social botomics: A systematic
  ensemble ml approach for explainable and multi-class bot detection,''
  \emph{Applied Sciences}, vol.~11, no.~21, 2021.

\bibitem{Ahmed_2013}
F.~Ahmed and M.~Abulaish, ``A generic statistical approach for spam detection
  in online social networks,'' \emph{Comput. Commun.}, vol.~36, no. 10-11,
  2013.

\bibitem{yang2020scalable}
K.-C. Yang, O.~Varol, P.-M. Hui, and F.~Menczer, ``Scalable and generalizable
  social bot detection through data selection,'' in \emph{Proceedings of the
  AAAI conference on artificial intelligence}, vol.~34, no.~01, 2020.

\bibitem{beskow2018using}
D.~M. Beskow and K.~M. Carley, ``Using random string classification to filter
  and annotate automated accounts,'' in \emph{{SBP-BRiMS}}.\hskip 1em plus
  0.5em minus 0.4em\relax Springer, 2018.

\bibitem{beskow2018bot}
------, ``Bot-hunter: a tiered approach to detecting \& characterizing
  automated activity on twitter,'' in \emph{{SBP-BRiMS}}, vol.~3, 2018.

\bibitem{beskow2018botconversations}
------, ``Bot conversations are different: leveraging network metrics for bot
  detection in twitter,'' in \emph{{ASONAM}}.\hskip 1em plus 0.5em minus
  0.4em\relax {ACM}, 2018.

\bibitem{Grimme_2018}
C.~Grimme, D.~Assenmacher, and L.~Adam, ``Changing perspectives: Is it
  sufficient to detect social bots?'' in \emph{{SCSM}}, vol. 10913.\hskip 1em
  plus 0.5em minus 0.4em\relax Springer, 2018.

\bibitem{YangCH_2013}
C.~Yang, R.~C. Harkreader, and G.~Gu, ``Empirical evaluation and new design for
  fighting evolving {Twitter} spammers,'' \emph{Trans. Inf. Forensics Secur.},
  vol.~8, no.~8, 2013.

\bibitem{magelinski2020graph}
T.~Magelinski, D.~Beskow, and K.~M. Carley, ``Graph-hist: Graph classification
  from latent feature histograms with application to bot detection,'' in
  \emph{Proceedings of the AAAI Conference on Artificial Intelligence},
  vol.~34, no.~04, 2020.

\bibitem{alhosseini2019detect}
S.~Ali~Alhosseini, R.~Bin~Tareaf, P.~Najafi, and C.~Meinel, ``Detect me if you
  can: Spam bot detection using inductive representation learning,'' in
  \emph{Companion Proceedings of The 2019 World Wide Web Conference}, 2019, pp.
  148--153.

\bibitem{feng2022heterogeneity}
S.~Feng, Z.~Tan, R.~Li, and M.~Luo, ``Heterogeneity-aware twitter bot detection
  with relational graph transformers,'' in \emph{Proceedings of the AAAI
  Conference on Artificial Intelligence}, vol.~36, no.~4, 2022.

\bibitem{kudugunta2018deep}
S.~Kudugunta and E.~Ferrara, ``Deep neural networks for bot detection,''
  \emph{Information Sciences}, vol. 467, 2018.

\bibitem{wei2019twitter}
F.~Wei and U.~T. Nguyen, ``Twitter bot detection using bidirectional long
  short-term memory neural networks and word embeddings,'' in
  \emph{TPS-ISA}.\hskip 1em plus 0.5em minus 0.4em\relax {IEEE}, 2019.

\bibitem{stanton2019gans}
G.~Stanton and A.~A. Irissappane, ``Gans for semi-supervised opinion spam
  detection,'' \emph{arXiv}, 2019.

\bibitem{feng2021satar}
S.~Feng, H.~Wan, N.~Wang, J.~Li, and M.~Luo, ``Satar: A self-supervised
  approach to twitter account representation learning and its application in
  bot detection,'' in \emph{{CIKM}}.\hskip 1em plus 0.5em minus 0.4em\relax
  {ACM}, 2021.

\bibitem{benigni2017online}
M.~C. Benigni, K.~Joseph, and K.~M. Carley, ``Online extremism and the
  communities that sustain it: Detecting the isis supporting community on
  twitter,'' \emph{PloS one}, vol.~12, no.~12, 2017.

\bibitem{wu2018bot}
W.~Wu, J.~Alvarez, C.~Liu, and H.-M. Sun, ``Bot detection using unsupervised
  machine learning,'' \emph{Microsystem Technologies}, vol.~24, no.~1, 2018.

\bibitem{koggalahewa2022unsupervised}
D.~Koggalahewa, Y.~Xu, and E.~Foo, ``An unsupervised method for social network
  spammer detection based on user information interests,'' \emph{J. Big Data},
  vol.~9, no.~1, 2022.

\bibitem{RuanWWJ16}
X.~Ruan, Z.~Wu, H.~Wang, and S.~Jajodia, ``Profiling online social behaviors
  for compromised account detection,'' \emph{Trans. Inf. Forensics Secur.},
  vol.~11, no.~1, 2016.

\bibitem{HDBSCAN_2017}
L.~McInnes, J.~Healy, and S.~Astels, ``{HDBSCAN}: Hierarchical density based
  clustering,'' \emph{J. Open Source Softw.}, vol.~2, no.~11, 2017.

\bibitem{DBSCAN_1996}
M.~Ester, H.~Kriegel, J.~Sander, and X.~Xu, ``A density-based algorithm for
  discovering clusters in large spatial databases with noise,'' in
  \emph{{KDD}}.\hskip 1em plus 0.5em minus 0.4em\relax {AAAI} Press, 1996.

\bibitem{macqueen_1967}
J.~MacQueen \emph{et~al.}, ``Some methods for classification and analysis of
  multivariate observations,'' in \emph{Proceedings of the fifth Berkeley
  symposium on mathematical statistics and probability}, vol.~1, no.~14, 1967.

\bibitem{lloyd_1982}
S.~Lloyd, ``Least squares quantization in pcm,'' \emph{Transactions on
  information theory}, vol.~28, no.~2, 1982.

\bibitem{chicco_2020}
D.~Chicco and G.~Jurman, ``The advantages of the matthews correlation
  coefficient (mcc) over f1 score and accuracy in binary classification
  evaluation,'' \emph{BMC genomics}, vol.~21, no.~1, 2020.

\bibitem{rahmah2016determination}
N.~Rahmah and I.~S. Sitanggang, ``Determination of optimal epsilon (eps) value
  on dbscan algorithm to clustering data on peatland hotspots in sumatra,'' in
  \emph{IOP conference series: earth and environmental science}, vol.~31,
  no.~1.\hskip 1em plus 0.5em minus 0.4em\relax IOP Publishing, 2016.

\bibitem{CresciDiPietro_2020}
S.~Cresci, R.~D. Pietro, M.~Petrocchi, A.~Spognardi, and M.~Tesconi, ``Emergent
  properties, models, and laws of behavioral similarities within groups of
  twitter users,'' \emph{Comput. Commun.}, vol. 150, 2020.

\end{thebibliography}

\end{document}